\let\origlabel\label 
\documentclass[lnbip]{svmultln}
\let\label\origlabel 
\usepackage{amsmath,amsfonts}
\usepackage{makeidx}  
\usepackage{relsize}
\usepackage{listings}
\usepackage{microtype}
\usepackage[hidelinks]{hyperref}
\usepackage[nameinlink]{cleveref}

\usepackage{float}
\usepackage[lofdepth,lotdepth]{subfig}
\usepackage{booktabs}
\usepackage{multirow}
\usepackage{tablefootnote}
\usepackage{todonotes}

\newfloat{lstfloat}{htbp}{lop}
\floatname{lstfloat}{Listing}
\crefalias{lstfloat}{listing}

\begin{document}
\mainmatter              
\title{A PUF-Based Approach for Copy Protection of Intellectual Property in Neural Network Models\thanks{The research reported in this paper has been funded by BMK, BMAW, and the State of Upper Austria in the frame of the COMET Module Dependable Production Environments with Software Security (DEPS) and the SCCH competence center INTEGRATE within the COMET - Competence Centers for Excellent Technologies Programme managed by Austrian Research Promotion Agency FFG.
The final publication is available at Springer via \url{https://doi.org/10.1007/978-3-031-56281-5_9}.}}
\titlerunning{A PUF-Based Approach for Copy Protection of IP in NN Models}  
%
\author{Daniel Dorfmeister, Flavio Ferrarotti, Bernhard Fischer,\\ Martin Schwandtner, Hannes Sochor}
\authorrunning{D. Dorfmeister et al.}   
%
%
\institute{Software Competence Center Hagenberg, Softwarepark 32a, 4232 Hagenberg, Austria\\
\texttt{\textless firstname\textgreater.\textless lastname\textgreater@scch.at}}

\maketitle              

\begin{abstract}        
More and more companies' Intellectual Property (IP) is being integrated into Neural Network (NN) models.
This IP has considerable value for companies and, therefore, requires adequate protection.
For example, an attacker might replicate a production machines' hardware and subsequently simply copy associated software and NN models onto the cloned hardware.
To make copying NN models onto cloned hardware infeasible, we present an approach to bind NN models---and thus also the IP contained within them---to their underlying hardware.
For this purpose, we link an NN model's weights, which are crucial for its operation, to unique and unclonable hardware properties by leveraging Physically Unclonable Functions (PUFs).
By doing so, sufficient accuracy can only be achieved using the target hardware to restore the original weights, rendering proper execution of the NN model on cloned hardware impossible.
We demonstrate that our approach accomplishes the desired degradation of accuracy on various NN models and outline possible future improvements.
\keywords {neural networks, intellectual property protection, physically unclonable functions, hardware-software binding}
\end{abstract}

\section{Introduction}


Neural Network (NN) models are increasingly deployed in all areas, including industry.
For instance, they are readily applied for quality optimisation and process automation in diverse industrial machines.
To make this possible, a considerable amount of resources, including time and money, are being invested in the development of NN models, which therefore increasingly incorporate the core Intellectual Property (IP) of companies.
In this context, it is important to ensure the privacy of the training data of the models, i.e., to prevent the extraction of the original data from the models. An important effort is indeed being made to deal with this problem~(see, e.g., \cite{Oliynyk2023}).

However, a rather neglected aspect is the fact that NN models can be easily copied and used without due authorisation.
This is known as software piracy and comprises the unauthorised copy, use, download, and distribution of software.
In this sense, a pre-trained NN model is not different to classical software, beyond the fact that it is learned instead of programmed.
We can differentiate among end-user, online and commercial piracy.
End-user piracy occurs when the software lacks copy protection and end-users illegally redistribute it in their private circle.
Online piracy is the distribution of pirated software online via a central server or a peer-to-peer file sharing platform.
Commercial piracy refers to an organisation that pirates software to gain a financial advantage, mainly by counterfeiting and then reselling it as original software or as part of a product that uses this software.
The main motivations for software piracy and their consequences for companies are well documented for a long time~\cite{doi:10.1287/mnsc.37.2.125}.
For all types of software piracy, the main problem is inadequate copy protection~\cite{softwarepiracy1}.

\begin{figure}[t]
    \centering
    \includegraphics[width=\columnwidth]{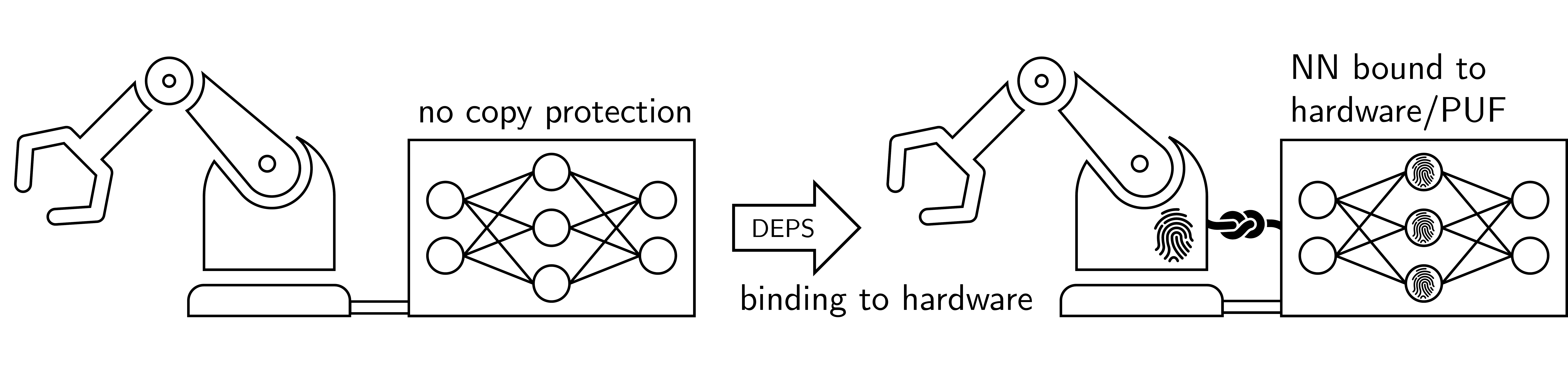}
    \caption{Copy Protection: A NN model tied to a target machine.}
    \label{fig:general-idea}
\end{figure}

In an industrial setting, the lack of copy protection built into NNs means that it is often not even necessary to know anything about a NN model, but it is enough to simply make a copy of it in order to profit at the expense of the IP owner.
In Germany alone, it is estimated that product or brand piracy accounts for an annual loss of 6.4 billion euros\footnote{VDMA Study Product Piracy 2022 (\url{https://www.vdma.org/documents/34570/51629660/VDMA+Study+Product+Piracy+2022_final.pdf})}.
The main effort when stealing IP from production machines is primarily in reverse engineering and cloning the hardware, as software and NN models can be simply copied.

In this paper, we propose a possible solution to the described problem.
Our solution differs from  classical copy protection mechanisms, usually  based on passwords or hardware dongles, as it is  well known that they are rather easy to circumvent, and can also be expensive if done properly.
We follow an approach based on binding a given NN model to a specific target machine so that its accuracy is reduced when copied (without authorisation) and used on a different machine.
We choose a protection approach based on making the copied NN model inaccurate instead of straight out unusable, as it can make the protection less obvious and thus more difficult to break for an attacker.
At the same time, this reduces the runtime overhead of authorised uses of the protected model.

In order to achieve the binding between a NN model and a target machine, i.e., the machine where the model is authorised to run on, we must be able to uniquely identify the underlying hardware. 
In the project DEPS\footnote{\url{https://deps.scch.at}} (short for \textsl{Dependable Production Environments with Software Security}) we are exploring different ways of achieving precisely this objective. In particular, a promising approach to software-hardware binding is the use of Physically Unclonable Functions (PUFs), a hardware-based security primitive.
This primitive arises from the fact that each circuit has unique physical properties resulting from unintended variations in the manufacturing process.
Consequently, these physical properties function as a digital fingerprint that cannot be easily cloned, which is the basis for PUFs.
PUFs can either use dedicated hardware or components already present in systems, such as DRAM.   


\begin{figure}[t]
    \centering
    \includegraphics[width=0.85\columnwidth]{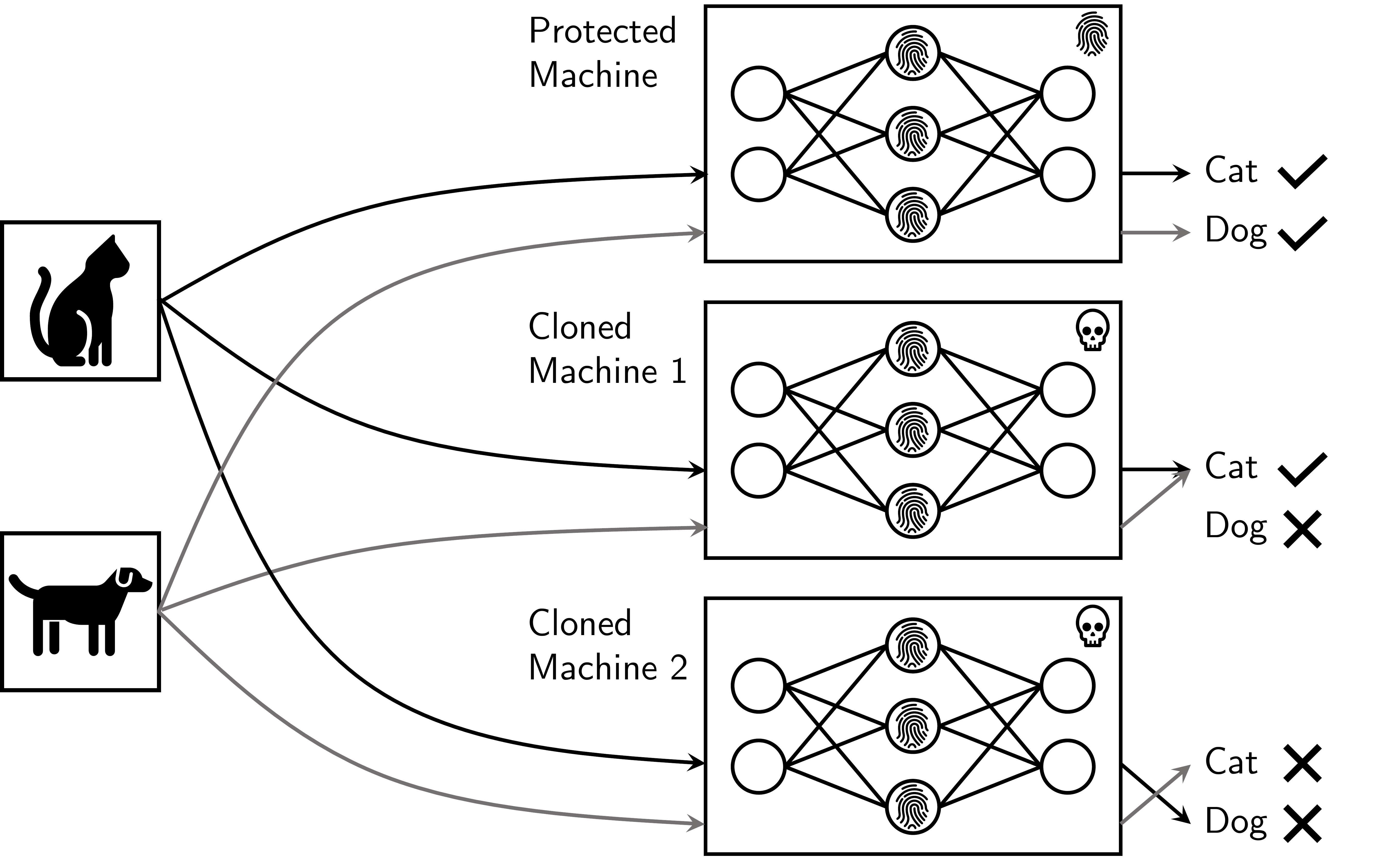}
    \caption{Behaviour of copy protected NN model in target vs. cloned machines.}
    \label{fig:how-it-works}
\end{figure}

\Cref{fig:general-idea} illustrates this general idea: We have an industrial machine, e.g., a robot arm with a neural network-based adaptive control method for trajectory tracking~\cite{robotArmExample}.
To protect it from being copied and used without authorisation, a fingerprint is taken from the industrial machine and tied to the neural network. 
More precisely, several weights of the NN model are bound to the fingerprint in such a way that the NN model can only provide sufficient accuracy on the basis of this fingerprint.
If such an industrial machine is now cloned and the protected NN model linked to the original hardware is copied to this cloned machine, the underlying fingerprint is now different and the NN model therefore no longer provides sufficient accuracy.

\Cref{fig:how-it-works} exemplifies the impact of our protection method on the accuracy of unauthorised copies of a NN model.
The protected target machine classifies the pictures of a dog and a cat correctly, while the two pirated copies  cannot verify the fingerprint (via the PUF) and thus classify them incorrectly.
Note that incorrect classifications might vary from one unauthorised copy to another.
This is because the fingerprint (PUF response) will differ between machines due to the unique physical characteristics of the individual cloned machines (see, e.g., cloned machines~1 and~2).
That is, if an adversary clones a machine and copies the protected NN model, then it will no longer work as reliable as on the target machine.


The paper is organised as follows:
In the next section, we briefly introduce the necessary background on NNs and PUFs, and fix the notation.
In \Cref{sec:threat}, we specify the threat model assumptions.
Our main contribution is condensed in \Cref{sec:protection_model,sec:poc,sec:evaluation,sec:discussion} where we, respectively, introduce our novel protection mechanism, a corresponding proof of concept strategy and implementation, associated experimental results, and discuss our findings and current limitations of the approach.
The final two Sections,~\ref{sec:relatedWork} and~\ref{sec:conclusion}, are reserved for related work and the conclusion, respectively.

\section{Preliminaries}





In this section, we provide the necessary background regarding neural network models and physically unclonable functions, and we fix the notation used throughout the paper.

\subsection{Neural Networks}\label{sec:NN}

There are many different kinds of (artificial) Neural Networks (NNs) such as feed forward (a.k.a. multilayer perceptrons), recurrent, and convolutional NNs.
All have in common that their architecture can be defined by some type of directed graph with a set of nodes $V$, set of edges $E \subseteq V \times V$ and functions $(a_v)_{v \in V}$, where for every node $v \in V$ (usually excluding the input nodes), $a_v:\mathbb{R} \rightarrow \mathbb{R}$ is a continuous function known as \emph{activation function} at $v$.
A concrete NN model associates this skeleton architecture with a \emph{weight} $w_e \in \mathbb{R}$ with every edge $e \in E$ and a \emph{bias} $b_v$ with every node $v \in V$.
NN models compute functions, but the way in which this is done varies depending on the type of NN used. 
A constant in this sense is that the computation takes place in the nodes and is propagated through the graph, using the activation functions, weights and biases.
The weight and biases are learned from data.
We are not concerned with the learning process itself but with protecting the IP produced by this process in the form of a pre-trained model.
Thus, it make sense to encrypt the weights in the pre-trained model so that an adversary cannot simply copy and use it without permission.



\subsection{Physically Unclonable Functions}
\label{sec:pufs}

A Physically Unclonable Function (PUF) $f$ is a function defined using unique physically properties of hardware that cannot be cloned.
Typically, it takes a binary string $b$ as input (challenge) and returns a possibly different binary string $b^\prime$ as output (response).
For each binary string $b$ in the domain of $f$, there is a corresponding $b^\prime = f(b)$, which is unique for the hardware of the target machine $M$.
More concretely, if $f$ is a PUF, $f(b)$ will be interpreted at run-time as $b'$ only if challenged on $M$.
Otherwise, the response value $f(b)$ will be arbitrary, and it can be assumed that $f(b) \neq b'$ for almost all machines $M'$ different from $M$. 

There is a variety of PUFs that can be used in the context of the protection proposed in this paper.
For instance, in our project DEPS, we are experimenting with DRAM PUFs~\cite{FischerBernhard2023DoaR}, where the PUF response $f(b) = b^\prime$ to a challenge $b$ is the result of applying the Rowhammer exploit to flip some bits of $b$, which is predictable in certain locations of the target DRAM.
Many other alternatives exist, e.g., arbiter PUFs, SRAM PUFs, ring oscillator PUFs, and optical PUFs~\cite{McGrath2019,10.1145/3591464}.
Any of these alternatives can be used for implementing the copy protection approach that we propose in this paper.
The choice depends on the available PUFs for the hardware where the trained NN model will be used, as well as the response time and level of strength provided by the available options.
This is application-specific and should be evaluated on a case by case basis.     




\section{Threat Model}\label{sec:threat}
In this paper, we only consider attacks that aim to remove the copy protection of the NN model, e.g., through reverse engineering.
We assume that an attacker can gain access to a model, e.g., via an update mechanism, backup of the machine, or download of the firmware from the machine.
We do not consider attacks on hardware such as side-channel attacks or other attack vectors, e.g., supply chain attacks.
This also means that we are currently looking at a mere static attacker.
This attacker has access to only the model and all the information it contains and can, therefore, not query the PUF and thus obtain Challenge-Response Pairs (CRPs).
However, we assume that our encryption method is public knowledge, as security through obscurity would be bad practice.
Both static and dynamic analysis are described below.
The focus of protection against dynamic analysis is future work and will be discussed in \Cref{sec:discussion}.



\subsection{Static Analysis}
Using static analysis, an attacker does not execute the software---including NN models---to be analysed but has access to the binary, potentially also to the source code of the software.
Thus, an attacker has access to all the information that is present while the program is not running.


For example, a piece of software uses an NN model and requires a key for decryption of the model's encrypted weights.
If the key is incorrect, the model's weights are not decrypted correctly and, therefore, the model does not achieve sufficient accuracy. 
However, if the encryption key is hidden somewhere within the program, an attacker may find this key through static analysis, decrypt the model and thus gain access to the model.

Attacks based on static analysis are possible as long as all data that is needed to successfully run a program or model is contained within themselves.
To counter such attacks, essential data must be inaccessible.
An example would be to prompt a third party, e.g., the user or a PUF for the decryption key so it does not have to be part of the program or model anymore.

\subsection{Dynamic Analysis}
Dynamic analysis is based on observing the behaviour and state of a program or NN at runtime.
This reaches from executing a program with various inputs and observing its behaviour from the outside by, e.g., dumping its memory, to employing a interactive disassembler.
Interactive disassemblers enable the reverse engineer to analyse a program at any time during its execution using breakpoints and enables inspection of memory at any state.

Dynamic analysis enables an attacker to use data only present at runtime in addition to information available from static analysis.
Suppose a cryptographic key is hidden successfully so that it cannot be retrieved by means of static analysis and it is also hidden at runtime.
The instructions of the program have to be decrypted at some point while executing the program.
However, an attacker can wait until each instruction has been decrypted at least once and can thus recover the full program. This also applies to NN. To successfully execute a NN, we must know the internal structure of the NN as well as its assigned weights. Whenever we need to perform some calculations using a weight, the correct, decrypted value must be provided.

In this work, we focus on protection against static analysis.
Therefore, our goal is to remove critical information from the model and make it only available at runtime.
Providing a sound method that prevents static analysis is a first step to secure NN models and already poses a significant challenge for attackers. 

\section{Copy Protection Method}
\label{sec:protection_model}

The goal of our copy protection method is to prevent unauthorised copy and use of pre-trained NN models.
We propose the use of PUFs (see~\Cref{sec:pufs}) to encrypt some of the weights $w_e$ associated with the edges $e \in E$ of an NN model, so that the model only works correctly on its target machine.
If used on a different machine---even on a clone of the target machine---, the model will drop its accuracy to levels that are determined by the number and selection of encrypted weights.
As we explained in~\Cref{sec:NN}, weights are learned from data and are a key IP asset contained in any pre-trained NN model, making them an ideal target for protection.
The reader might ask why we encrypt just some of the weights and not all of them.
The answer is that this way our method can limit the performance overhead caused by decryption and, as shown in this paper, still provide an adequate level of protection.   

More concretely, given a NN model $N$ with set of edges $E$ and corresponding weights $w_e$ for every $e \in E$, plus a PUF $f$ with domain $D_f$ for a target machine $M$, we propose a protection method based on binding $N$ to $M$ (so that $N$ only works correctly on $M$) via encryption/decryption of a subset of the weights of $N$ using $f$.
Before the encryption process starts, we need to decide the number $n$ of weights from $N$ that we want to encrypt.
This number depends on various factors that are application specific.
In particular, these are the desired drop in accuracy of $N$ when used on unauthorised hardware and the performance overhead when used on the authorised hardware.
We analyse this issue over some concrete NN models in \Cref{sec:evaluation}. 

For the encryption procedure, we adapt the well known one-time key encryption mechanics in the special form of Vernam~\cite[Section~13.2]{Biskup09}, using the PUF $f$ (by selecting a challenge at random) to generate the cipher key.
We work with the binary representation of the real-valued weights.
Thus, we use $\mathit{toBin}(w_e) = b_e$ and $\mathit{toFloat}(b_e) = w_e$ to denote the binary representation $b_e$ of weight $w_e$ and its inverse function, respectively.
We assume that the binary representation of a weight is of length $m$, i.e., $|\mathit{toBin}(w_e)| = m$ for all weight $w_e$ of $N$.
We further assume that the PUF $f$ is challenged with binary strings of length $m$ and responds with binary strings of the same length. 

The method to copy protect $N$ can be outlined as follows:
\begin{itemize}
    \item The \emph{weight selection} algorithm chooses a subset $S \subseteq E$ of edges of size $|S| = n$. 
    \item For each weight $w_e$ with $e \in S$, the \emph{key generation} algorithm chooses a (challenge) $c_e \in D_f$ randomly.
    \item For each weight $w_e$ with $e \in S$, the \emph{encryption} algorithms handles the binary representation $\mathit{toBin}(w_e) = (w_{e1}, \ldots, w_{em})$ (the plaintext) and the PUF response $f(c_e) = (k_{e1}, \ldots k_{em})$  (the cipher key) as streams, and uses each corresponding pair of a plaintext bit $w_{ei}$ and a cipher key bit $c_{ei}$ as input for a XOR operation, yielding a ciphertext bit $p_{ei} = k_{ei} \oplus w_{ei}$. 
    \item For each ciphertext (encrypted weight)  $p_e = (p_{e1}, \ldots, p_{em})$, the \emph{decryption} algorithm obtains the corresponding key using the PUF response $f(c_e) = (k_{e1}, \ldots k_{em})$ and treats each corresponding pair of a ciphertext bit $p_{ei}$ and a cipher key bit $k_{ei}$ as input for a XOR operation. Since \[k_{ei} \oplus p_{ei} = k_{ei} \oplus (k_{ei} \oplus w_{ei}) = (k_{ei} \oplus k_{ei}) \oplus w_{ei}) = 0 \oplus w_{ei} = w_{ei},\] the XOR operation yields the original bit in the binary representation of $w_e$ (i.e., the original plaintext bit). Finally, $\mathit{toFloat}(w_{e1}, \ldots, w_{em}) = w_e$.           
\end{itemize}

Note that the use of the PUF $f$ to retrieve the key means that decryption of the weights is correct only if the response to the challenges chosen during encryption is the one given by the PUF $f$ on the target machine $M$.
In case a hardware component the PUF is based on needs to be replaced, the NN model must be decrypted first an re-encrypted using the new hardware.
Alternatively,---especially if the hardware component failed---the original, unencrypted model can be encrypted for the new hardware and re-deployed to the altered target machine.

\section{Proof of Concept}
\label{sec:poc}

As a Proof of Concept (PoC), we implemented the copy protection method for NNs described in the previous section using Python~3.10 and TensorFlow~2.14.
The pseudocode in \Cref{lst:setup} shows how our PoC implements the weight selection, key generation and encryption.

\begin{lstfloat}
\begin{lstlisting}
def encrypt_model(model, layer_id, pct, data):
  print "Accuracy = " + model.evaluate(data) # = 0.97
  chosen_weights :=
    choose_weights(model.layers[layer_id].weights, pct) 
  
  foreach weight_id in chosen_weights:
    w := model.layers[layer_id].weights[weight_id]
    choose challenge in domain(puf): # randomly
      p := puf(challenge) xor w
      model.layers[layer_id].weights[weight_id] := p
      helper[weight_id] := challenge

  save(helper, layer_id)
  print "Accuracy = " + model.evaluate(data) # e.g., = 0.75
\end{lstlisting}
    \caption{Encrypt model's weights (before NN deployment).}
    \label{lst:setup}
\end{lstfloat}

The algorithm \texttt{encrypt\_model} takes the following parameters: the NN model (\texttt{model}), the layer that we want to encrypt (\texttt{layer\_id}), the percentage of weights of a given layer that must be encrypted (\texttt{pct}), and input data (\texttt{data}) to test the accuracy of the model with both the original and modified weights after the encryption.
Note that we use the layer for evaluation purposes since it is a known fact that certain layers are more significant than others.

Our PoC starts by testing the accuracy of the model for classifying the inputs in \texttt{data} (line~2), which is repeated once the model is encrypted (line~14) to test the  accuracy of the resulting model with modified (encrypted) weights.
This way, we can determine whether the given percentage of encrypted weights is sufficient to degrade the model's responses, rendering unauthorised copies of the model useless.  

The subset of weights to be encrypted (\texttt{chosen\_weights}) is randomly chosen from weights in the selected layer of the model, where the cardinality of \texttt{chosen\_weights} amounts to the percentage \texttt{pct} of all weights in that layer (lines~3--4).
By increasing \texttt{pct}, we can increase the strength of our protection, at the cost of runtime overhead while encrypting and decrypting the model.

Next (lines~6--11), the key generation and encryption is performed for each of the chosen weights.
Note that we keep track of the corresponding \texttt{challenge} for each \texttt{weight\_id} using the \texttt{helper} vector.
This is needed to recover the key via the \texttt{puf} function when the encrypted weight \texttt{p} needs to be decrypted, which only works correctly on the target machine.




Currently, we use a simulated XOR arbiter PUF~\cite{xorarbiterpuf}, implemented in the pypuf library\footnote{\url{https://pypuf.readthedocs.io}}.
This allows our PoC to work on any PC for demonstration purposes (as long as pypuf is installed).
We can decide which machine's PUF responses we want to simulate by providing a seed to pypuf.
In practice, we should of course use an actual PUF for the target machine, instead of a simulated one.
Then, the model should be encrypted on the target machine using our algorithm in~\Cref{lst:setup}, or alternatively by modifying the algorithm to use a database that maps PUF challenges to the target machine's responses.


\begin{lstfloat}
\begin{lstlisting}
def decrypt_model(model, data):
  helper, layer_id := load()
  
  foreach (weight_id, challenge) in helper:
    p := model.layers[layer_id].weights[weight_id]
    w := puf(challenge) xor p
    model.layers[layer_id].weights[weight_id] := w
  
  print "Accuracy = " + model.evaluate(data)
  # Accuracy = 0.97 (on target machine)
  # Accuracy < 0.97 (on cloned/different machine)
\end{lstlisting}
    \caption{Decrypt model's weights (when executed on the target machine).}
    \label{lst:runtime}
\end{lstfloat}

In \Cref{lst:runtime}, we describe our corresponding PoC approach for decryption of weights of a given NN model.
This procedure results in the correct model if, and only if, the PUF's responses to the required challenges coincide with the responses used during encryption.
In turn, this should only happen in practice if the decryption algorithm is run on the target machine.
For added security against static attacks, we propose to apply the decryption in memory at loading time, i.e., right before the model is used.  

The \texttt{decrypt\_model} algorithm takes the following parameters: the NN model to be decrypted (\texttt{model}), and the input data used to measure the accuracy of the decrypted model (\texttt{data}).
It also needs the \texttt{helper} vector created during encryption, which also contains the IDs of the \texttt{chosen\_weights} as keys, as well as the used \texttt{layer\_id}.
This information is loaded at the beginning of the decryption process (line~2). 
The decryption procedure is done using the approach explained in the previous section.
It simply uses the PUF responses and the encrypted weights to retrieve the original weights via a bitwise XOR operation (lines~5--8).

Note that we test the actual accuracy of the decrypted model at the end of the decryption process.
If executed on the target machine (simulated in our PoC by using the same seed for the PUF simulator as for encryption), the accuracy should match the original accuracy of the model (i.e., before encryption).
Otherwise, the accuracy of the NN model will be necessary lower, as planned.
The expected incorrect PUF response in the latter case means that the decrypted weight does not match the original weight.


\section{Evaluation}
\label{sec:evaluation}



We choose four different NN models to evaluate our experiments.
We train these models ourselves following TensorFlow tutorials:
Two of our models classify images, one is based on the MNIST dataset\footnote{\url{https://www.tensorflow.org/datasets/catalog/mnist}} to detect handwritten digits, and the other one on the Fashion MNIST dataset~\cite{fashion-mnist} to detect ten categories of clothing.
The other models recognise eight different predefined speech commands~\cite{speechcommandsv2}, and positive or negative sentiment in IMDB reviews~\cite{imdb-reviews}.
We encrypt a dense layer of each image classification model.
For the latter two models, we encrypt one convolutional and one recurrent layer, respectively.
For an overview of all experiments, the models we use, and the total number of weights in the layers we encrypt, see \Cref{tab:experiments}.


\begin{table}[t]
    \centering
    \begin{tabular}{llllr}
        Exp. & Model Type & Classes                                  & Layer         & \# Weights \\ \toprule
        (a)  & Image Classification\tablefootnote{\url{https://www.tensorflow.org/datasets/keras_example}}                  
                          & 10 (digits 0--9)                         & Dense         & $100\,352$ \\
        (b)  & Image Classification\tablefootnote{\url{https://www.tensorflow.org/tutorials/keras/classification}}
                          & 10 (clothing categories)                 & Dense         & $100\,352$ \\ \midrule
        (c)  & Audio Recognition\tablefootnote{\url{https://www.tensorflow.org/tutorials/audio/simple_audio}}
                          & 8 (audio commands)                       & CNN           &  $18\,430$ \\
        (d)  & Text Recognition\tablefootnote{\url{https://www.tensorflow.org/text/tutorials/text_classification_rnn}}    
                          & 2 (pos./neg. sentiment)                  & RNN           &  $16\,384$ \\
    \end{tabular}
    \caption{Overview of the models/layers used in the experiments in \Cref{fig:evaluation}.}
    \label{tab:experiments}
\end{table}

\subsection{Accuracy}

\begin{figure}[htp]
    \centering
    \subfloat[Image Classific. (Digits, Dense Layer)]{
        \includegraphics[width=0.5\textwidth]{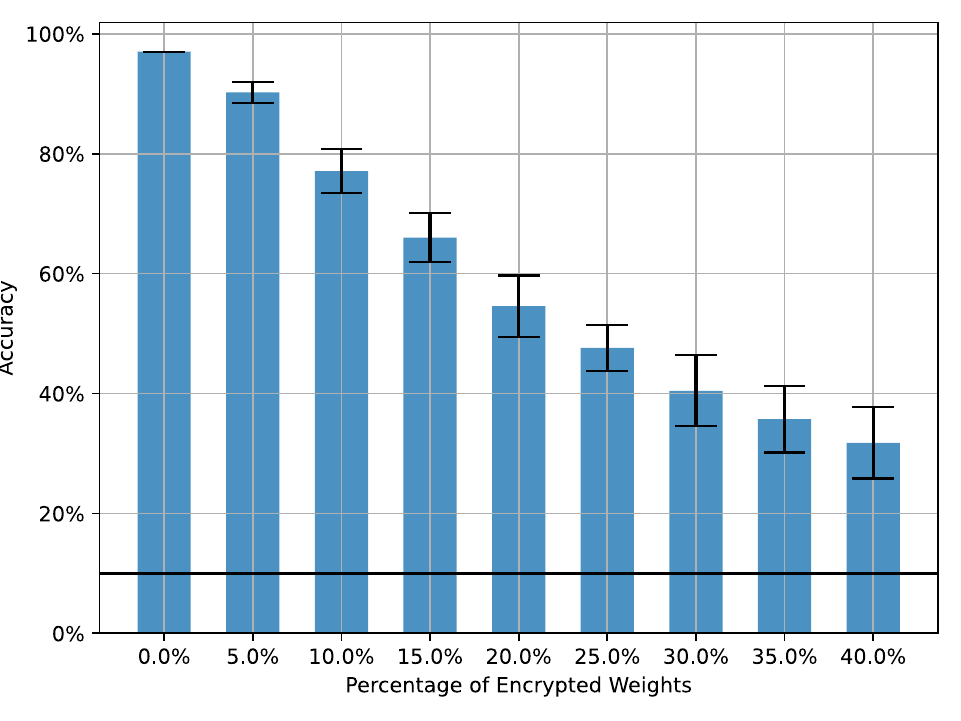}
        \label{fig:subfig1}}
    \subfloat[Image Classific. (Clothing, Dense Layer)]{
        \includegraphics[width=0.5\textwidth]{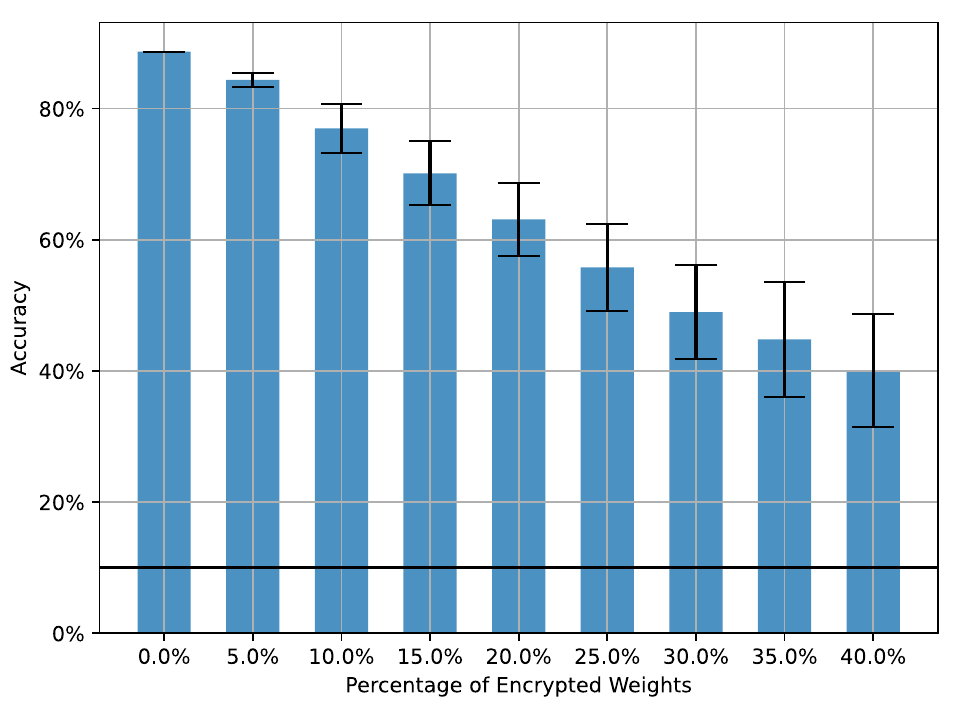}
        \label{fig:subfig2}}
    \qquad
    \subfloat[Audio Recognition (CNN Layer)]{
        \includegraphics[width=0.5\textwidth]{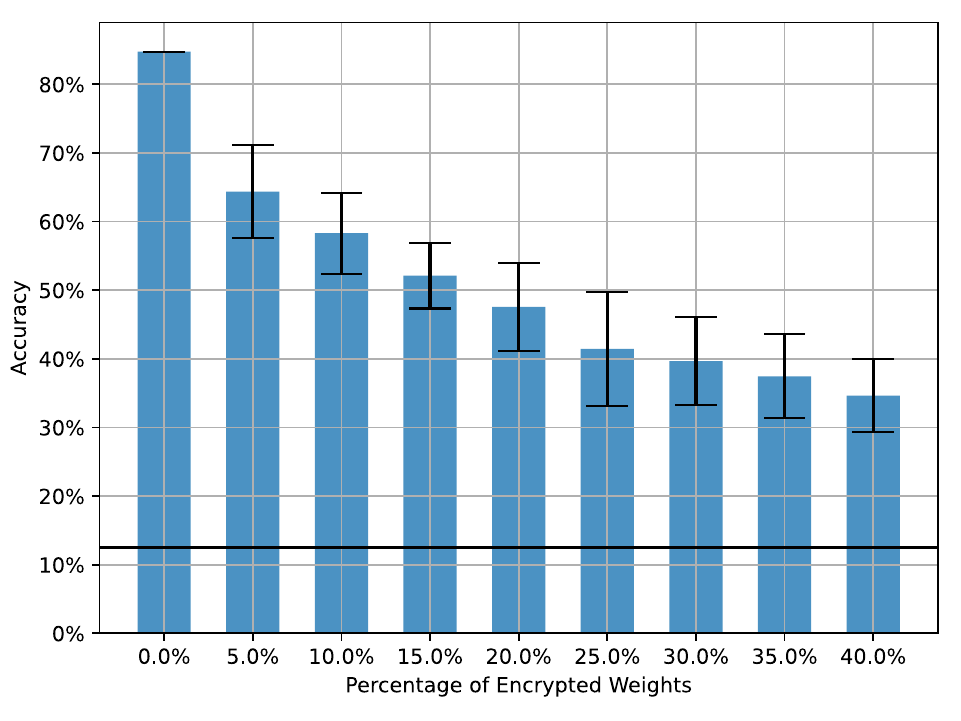}
        \label{fig:subfig4}}
    \subfloat[Text Classification (RNN Layer)]{
        \includegraphics[width=0.5\textwidth]{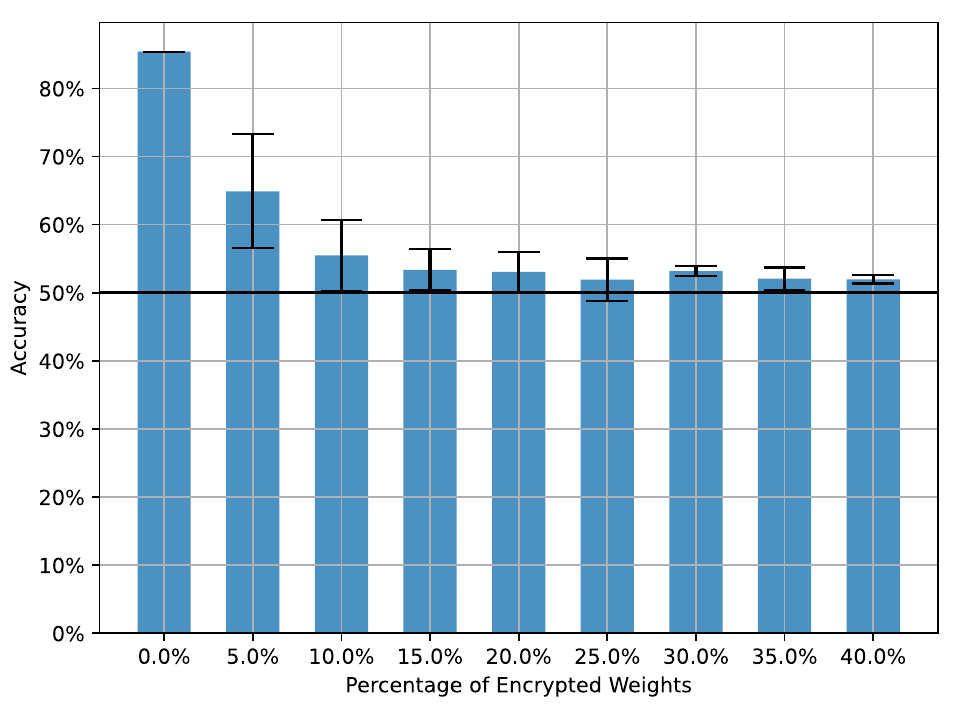}
        \label{fig:subfig5}}
    \caption{Mean accuracy drop (and standard deviation) for the models described in \Cref{tab:experiments} depending on percentage of encrypted weights for 10 randomly chosen sets of weights each. For comparison, the black horizontal lines represent random classifiers.}
    \label{fig:evaluation}
\end{figure}

To show the degradation of accuracy in our models due to encryption, we randomly select ten sets of weights containing 40~\% of the weights per layer to be encrypted.
For each of these, we also collect data for subsets including fewer weights at 5 percentage point decrements.
For comparison, we also measure the accuracy at 0~\% encryption, i.e., an unencrypted model.
\Cref{fig:evaluation} shows the mean of the results of our experiments as well as their standard deviation.
Note that these results are only valid for executing the encrypted models without trying to decrypt them.

For all experiments, the accuracy of the models drops significantly even at 5~\% encrypted weights and approximates a random classifier (symbolised by the black horizontal line in \Cref{fig:evaluation}) at higher percentages of encrypted weights.
For example, in \Cref{fig:subfig5} the text classification model's accuracy drops to the value expected for a random classifier at just 10--20~\%, making the encryption of additional weights unnecessary.
By selecting the weights to encrypt carefully, focusing on weights with the most impact, we could decrease the number of weights that must be encrypted in order to achieve the desired level of protection.
For more details on this, see \Cref{sec:relatedWork}.

\begin{figure}[ht]
    \centering
    \includegraphics[width=.8\columnwidth]{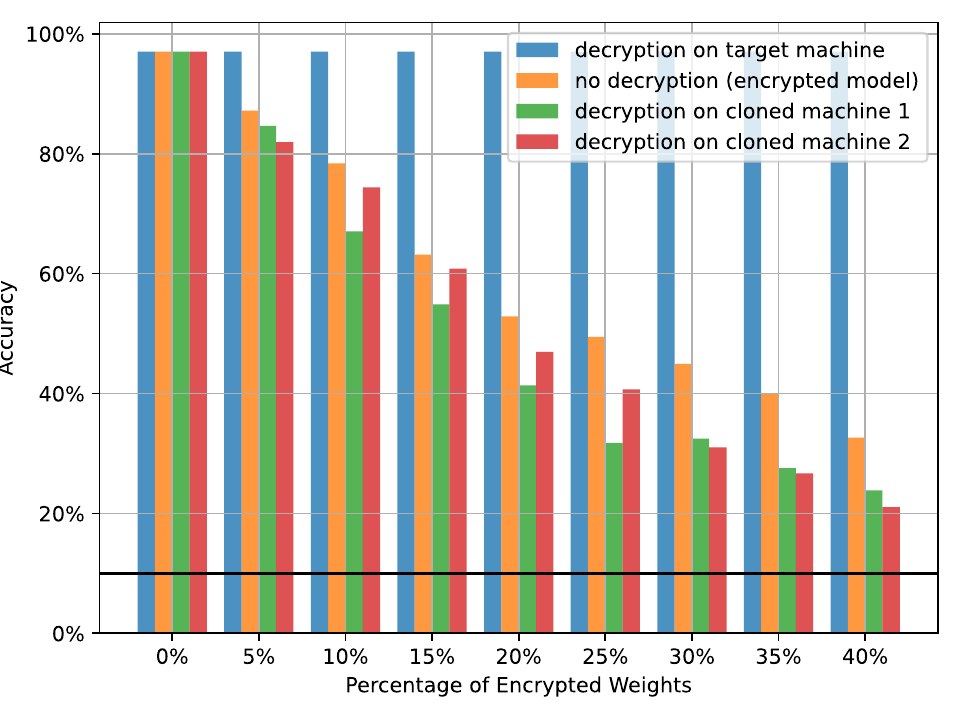}
    \caption{Comparison of the accuracy  of the image classifier from \Cref{fig:subfig1} at different levels of encryption when leaving it encrypted and decrypting it on various machines, i.e., the target machine and two cloned machines. The black horizontal line represents a random classifier.}
    \label{fig:decryption}
\end{figure}

\Cref{fig:decryption} shows the accuracy of the encrypted model from experiment (a) when used without decryption compared to decryption on various machines.
When we decrypt the model on its target machine, we achieve perfect decryption and restore the model's original accuracy.
In case the encrypted model is used without decryption, i.e., by extracting it from our software and thus circumventing the execution of the decryption algorithm, the accuracy is lowered (cf. \Cref{fig:subfig1}).
Decryption on machines different from the target machine, e.g., cloned machines, potentially further degrades the accuracy of the model.
On these machines, the PUF responses differ from those on the target machine and thus, decryption of the encrypted weights fails.
Effectively, the weights are encrypted once more, potentially lowering the model's accuracy even further.

\subsection{Size Impact}

The size of the actual NN model does not change during encryption.
Nevertheless, we need to store helper data containing the IDs of the encrypted weights and their respective challenges for the PUF.
With our PoC implementation, the size of this data structure increases by 12 bytes per additional encrypted weight.
For example, the helper data for the model we use in experiment (a) at 20~\% encrypted weights, which yields a significant degradation of the model, requires approximately 2.3~MiB of helper data, which could be further reduced by using a more efficient way to identify the weights.
The helper data must only be loaded from disk during decryption, and it is not required to load this data into main memory at once.
Thus, even on embedded devices, the additional data should not hinder the application of our approach.


\section{Discussion}
\label{sec:discussion}

The used one-time key encryption mechanics in the form of Vernam that we apply in this paper achieves \emph{perfect} security in a  precise probability-theoretic sense~\cite[Theorem~12.2]{Biskup09}.
This means that it is  qualified to the best possible extent regarding the secrecy property and the efficiency property of the encryption and decryption algorithms.
However, this only holds if the key is (a) truly randomly selected, (b) used only once, and (c) as long as the plaintext. 
As usual in practice, there are trade-offs in our adaptation to the case of protecting NN models.

In theory, we can meet properties~(a) and~(c) if the PUF $f$ is injective with domain and range $\{0, 1\}^m$, where $m$ is the length of the binary representation of the weights of the NN model $N$.
In practice, we most probably need to replace the “truly random” cipher key by a pseudorandom one, since PUFs are usually not injective and shorter than the required length $m$.
The same applies, for instance, to the Vigen\`ere encryption mechanism~\cite[Section~13.3]{Biskup09}, this inevitably results in a less secure encryption.
Property (b), on the other hand, can clearly be met, as long as the key generation mechanism does not impose unacceptable overheads.

As our PoC implementation uses a simulated PUF and is not optimised for performance, we did not conduct a performance evaluation.
Nevertheless, it is clear that the performance will be heavily influenced by the PUF response time as well as the number of encrypted weights.

It is also apparent that there is a trade-off between performance overhead and security level.
This is determined by the desired drop in accuracy on the one hand and the type of encryption on the other.
If the model is only decrypted once after loading (as per our PoC implementation), then the PUF response time does not affect the runtime, only the loading time is affected.
If we use a more secure approach instead where each individual weight is decrypted on demand for each processed input, and then encrypted again immediately after use (possibly with a new key), then this will clearly have a negative impact in the NN model's performance.
At the same time, the resulting protection would have better resistance to sophisticated (dynamic) reverse engineering attacks. 

In our PoC, we select weights at random and show that the encryption of around 20~\% of the weights is enough to obtain a sufficient degradation in the accuracy of the NN models, and thus being able to apply the proposed copy protection against static attacks.
To offer protection against dynamic attacks as outlined in the previous paragraph, but without affecting the performance significantly, one could potentially encrypt/decrypt a substantially smaller number of weight for each input processed by the NN model.
This can still degrade the accuracy of the NN model enough (see, e.g., \cite{frantar2023sparsegpt,Ruospo2022}).
We plan to investigate this alternative in future research. 

A final point that should be noted is that if attackers can somehow query the PUF on the target machine to get the needed responses, then they can obviously decrypt the network.
There are, however, known ways to make PUFs resistant to this type of attacks (see, e.g.,~\cite{McGrath2019,10.1145/3591464}).
Additionally, we could use obfuscation techniques so that an attacker cannot easily identify where each PUF response is used.
There is also the possibility of considering Trusted Execution Environments (TEEs) for key operations or homomorphic encryption.
Again, this needs to be investigated in future work and will necessary involve a trade-off between security and performance.

\section{Related Work}\label{sec:relatedWork}
Protecting the IP in NN models is of course not a completely new idea. Indeed, several alternative methods already exist.
In this section, we will discuss the most relevant work in this sense. 

We can identify two main approaches depending on how IP protection is applied: (i) by means of different obfuscation techniques~\cite{9417217, 8203897} that make it sufficiently harder for an attacker to recover the original model, and (ii) by using cryptography~\cite{pan2022devicebind, 8567421, 9171467} to either encrypt the whole model, individual layers or, as in our case, individual weights.
In the remainder of this section, we discuss each of these related techniques. 

The Goldstein et al.~\cite{9417217} have shown that a good level of protection may be achieved by applying various alterations to the structure of convolutional filters in Deep Convolutional Neural Networks (DCNNs).
To revert the applied obfuscation, a secret key is needed.
If an incorrect key is provided, the resulting model will have a significantly lower accuracy than the original model.

Rakin et al.~\cite{rakin2019bitflip} explore the effects of applying bit flips induced by Rowhammer attacks to a neural network.
The authors use these bit flips to attack a model and render it useless for other users.
While this idea is not directly related to our work, Zhao et al.~\cite{8203897} use a similar technique to protect neural networks.
They propose to calculate an error mask based on DRAM restore values that pose as a unique physical property of individual DRAM modules.
As such, the authors are able to create an error mask that is bound to a specific DRAM module.
They proceed to use this error mask while training a neural network.
The resulting model is then bound to the DRAM module that will yield the correct error mask.
If executed on another DRAM module, the accuracy of the model is significantly lower than the accuracy of the original model.
The key  difference to our approach is that their model needs to be individually trained for each piece of hardware it runs on while our work can be applied to any pre-trained model.

Cai et al.~\cite{9171467} apply encryption to only a small portion of the weights.
They apply the proposed encryption using a pre-calculated set of secret keys that are stored in a separate key storage.
The authors show that encrypting only a small portion of the weights is enough to achieve a sufficient level of protection.
In contrast to our work, they rely on a set of keys that, if known to an attacker, can be used to run the network at any arbitrary machine whereas our approach is bound to specific hardware by using a PUF to generate secret keys while encrypting and decrypting the model.

Making use of FPGAs to accelerate the execution time of neural networks is a well known method.
Guo et al.~\cite{8567421} propose to directly integrate their protection into such an accelerator FPGA.
They make use of a PUF that only yields the correct results when executed on the correct FPGA.
While their method seems very promising, particularly regarding the low performance overhead associated with the proposed protection, the solution is specifically tailored to convolutional NNs running on FPGA hardware.

Finally, Pan et al.~\cite{pan2022devicebind} use an Anderson PUF to incrementally encrypt layers of a NN to efficiently bind it to specific hardware.
They focus on analysing the effectiveness of the protection to counteract fine-tuning attacks.
This is a type of attack that tries to approximate the original model from the obfuscated one by making use of a fraction of the original training dataset.
A key difference to our approach is that they aim to render the behaviour of the protected model equivalent to a random classifier.
In contrast, our work has identified that obfuscating only a small fraction of the weights is enough to achieve a good level of protection against software piracy.
Since we can limit the loss in accuracy to a few percentage points, in theory we  should be able to make our approach more stealthy, which would in turn give us an advantage w.r.t. the fine-tuning attacks considered in their work.
The conjecture is that then it is not immediately obvious for an attacker how the encrypted NN model produces the drop in performance.
This still needs to be confirmed by future research, though.

\section{Conclusion}\label{sec:conclusion}

In this work, we presented a method to protect intellectual property in neural network models from piracy.
The main advantage of our approach is that NN models cannot simply be copied to replicated machines, thus requiring reverse engineering not just for the hardware but also the NN models.
We evaluated our method using different NN models and showed that they can be bound to unique hardware properties in such a way that copying an NN model to another machine renders it useless.
Furthermore, this complicates a static analysis, since without knowledge of the properties of the target hardware, i.e., PUF responses, this mechanism cannot simply be removed.

A further step and planned future work is to make this protection more robust against dynamic analysis.
The current PoC implements the decryption after loading, i.e., the entire model is decrypted and then stored decrypted in main memory.
A possible improvement in this respect would be to decrypt only individual weights that are currently being used and then encrypt them again, or to integrate this mechanism into a trusted execution environment.
In addition, it is also necessary to analyse and optimise performance and subsequently achieve a good trade-off between performance and security, i.e., the number of weights to be encrypted.

%
%
\bibliographystyle{splncs04}
\bibliography{references}

\end{document}